\documentclass[sigconf]{acmart}

\AtBeginDocument{%
  }

\setcopyright{acmcopyright}
\copyrightyear{}
\acmYear{}
\acmDOI{}

\acmConference[]{}{}{}
\acmPrice{}
\acmISBN{}

\begin{document}

\title{Making an agent's trust stable in a series of success and failure tasks through empathy}

\author{Takahiro Tsumura}
\email{takahiro-gs@nii.ac.jp}
\orcid{0000-0002-3145-3120}
\affiliation{%
  \institution{The Graduate University for Advanced Studies, SOKENDAI}
  \institution{National Institute of Informatics}
  \streetaddress{2-1-2 Hitotsubashi}
  \city{Chiyoda-ku}
  \state{Tokyo}
  \country{Japan}
  \postcode{101-8430}
}

\author{Seiji Yamada}
\email{takahiro-gs@nii.ac.jp}
\orcid{0000-0002-5907-7382}
\affiliation{%
  \institution{National Institute of Informatics}
  \institution{The Graduate University for Advanced Studies, SOKENDAI}
  \streetaddress{2-1-2 Hitotsubashi}
  \city{Chiyoda-ku}
  \state{Tokyo}
  \country{Japan}
  \postcode{101-8430}
}


\begin{abstract}
As AI technology develops, trust in AI agents is becoming more important for more AI applications in human society. 
Possible ways to improve the trust relationship include empathy, success-failure series, and capability (performance). 
Appropriate trust is less likely to cause deviations between actual and ideal performance. 
In this study, we focus on the agent's empathy and success-failure series to increase trust in AI agents. 
We experimentally examine the effect of empathy from agent to person on changes in trust over time. 
The experiment was conducted with a two-factor mixed design: empathy (available, not available) and success-failure series (phase 1 to phase 5). 
An analysis of variance (ANOVA) was conducted using data from 198 participants. 
The results showed an interaction between the empathy factor and the success-failure series factor, with trust in the agent stabilizing when empathy was present. 
This result supports our hypothesis. 
This study shows that designing AI agents to be empathetic is an important factor for trust and helps humans build appropriate trust relationships with AI agents.
\end{abstract}




\maketitle

\section{Introduction}
Humans live in society and use a variety of tools, but AI is sometimes relied upon more than humans. 
Indeed, today's AI issues concern the trustworthiness and ethical use of AI technology. 
Ryan \cite{Ryan20} focused on trustworthiness and discussed AI ethics and the problem of people anthropomorphizing AI. 
He determined that even complex machines that use AI should not be viewed as trustworthy. Instead, he suggested that organizations that use AI and the individuals within those organizations should ensure that they are trustworthy. 
The work of Hallamaa and Kalliokoski \cite{Hallamaa22} also discussed AI ethics in depth from an applied ethics perspective. 
Kaplan et al. \cite{Kaplan23} aimed to identify important factors that predict trust in AI and examined three predictive categories and subcategories of human characteristics and abilities, AI performance and attributes, and contextual challenges from data from 65 articles. 
All of the categories examined were significant predictors of trust in AI. 
The more AI is used in human society, the more trust in AI is discussed, and it is an important issue that failure to establish appropriate trust relationships can lead to overconfidence and distrust of AI agents, which in turn reduces task performance.
\\ \indent
In recent years, research on trust agents has also received attention. 
In a study of trustworthy AI agents, Maehigashi et al. \cite{Maehigashi22-1} investigated how beeps emitted by a social robot with anthropomorphic physicality affect human trust in the robot. 
They found that (1) sounds just prior to a higher performance increased trust when the robot performed correctly, and (2) sounds just prior to a lower performance decreased trust to a greater extent when the robot performed inaccurately. 
To determine how anthropomorphic physicality affects human trust in agents, Maehigashi et al. \cite{Maehigashi22-2} also investigated whether human trust in social robots with anthropomorphic physicality is similar to trust in AI agents and humans. 
Also, they investigated whether trust in social robots is similar to trust in AI agents and humans. 
The results showed that trust in social robots was basically not similar to trust in AI agents or humans, and was entrenched between the two.
\\ \indent
Along with trust, we often empathize with artifacts. Humans are known to have a tendency to treat artifacts as if they were humans in the media equation \cite{Reeves96}. 
However, some humans do not accept these agents \cite{Nomura06,Nomura08,Nomura16}. 
Empathy is closely related to trust. As agents permeate society in the future, it is hoped that they will have elements that are acceptable to humans.
\\ \indent
Omdahl \cite{Omdahl95} roughly classifies empathy into three types: (1) affective empathy, which is an emotional response to the emotional state of others, (2) cognitive understanding of the emotional state of others, which is defined as cognitive empathy, and (3) empathy including the above two. 
Preston and de Waal \cite{Preston02} suggested that at the heart of the empathic response was a mechanism that allowed the observer to access the subjective emotional state of the target. 
They defined the perception-action model (PAM) and unified the different perspectives in empathy. 
They defined empathy as three types: (a) sharing or being influenced by the emotional state of others, (b) assessing the reasons for the emotional state, and (c) having the ability to identify and incorporate other perspectives. 
\\ \indent
In this study, we design the agent to have the ability to empathize with people so that trust in the agent can remain appropriate. 
Therefore, empathy in this study is about the agent's ability to empathize with people. 
To investigate the possible influence of empathy on changes in trust over time, we also investigate trust in the agent for each phase in a total of five phases.

\section{Related work}
There have been a number of recent papers on trust in AI, and we summarize some of the previous studies that have studied AI performance and relationships in particular. 
Kumar et al. \cite{Kumar2023} noted that new technological advances with the use of AI in medicine have not only raised concerns about public trust and ethics, but have also generated much debate about its integration into medicine. 
They reviewed current research investigating how to apply AI methods to create smart predictive maintenance. 
Noting that lack of trust is one of the main obstacles standing in the way of taking full advantage of AI, Gillath et al. \cite{Gillath21} focused on increasing trust in AI through emotional means. Specifically, they tested the association between attachment style, an individual difference that describes how people feel, think, and act in relationships, and trust in AI. 
Results showed that increasing attachment insecurity decreased trust, while increasing attachment security increased trust in AI. 
Focusing on clinicians as the primary users of AI systems in healthcare, Asan et al. \cite{Asan20} presented the factors that shape trust between clinicians and AI. 
They focused on key trust-related issues to consider when developing AI systems for clinical use.
\\ \indent
Maehigashi \cite{Maehigashi22-3} experimentally investigated the nature of human trust in communication robots compared to trust in other people and AI systems. 
Results showed that trust in robots in computational tasks that yield a single solution is essentially similar to that in AI systems, and partially similar to trust in others in emotion recognition tasks that allow multiple interpretations.
Okamura and Yamada \cite{Okamura20-1} proposed a method for adaptive trust calibration that consists of a framework for detecting inappropriate calibration conditions by monitoring the user's trust behavior and cognitive cues, called ``trust calibration cues," that prompt the user to resume trust calibration. 
Okamura and Yamada \cite{Okamura20-2} focused their research on trust alignment as a way to detect and mitigate inappropriate trust alignment, and they addressed these research questions using a behavior-based approach to understanding calibration status. 
The results demonstrate that adaptive presentation of trust calibration cues can facilitate trust adjustments more effectively than traditional system transparency approaches.
\\ \indent
Kahr et al. \cite{Kahr23} focused on trust in appropriate AI systems and how trust evolves over time in human-AI interaction scenarios. 
Results showed significantly higher trust in the high accuracy model, with behavioral trust not decreasing and subjective trust increasing significantly with higher accuracy.
Ma et al. \cite{Ma23} proposed promoting appropriate human trust on the basis of the correctness likelihood of both sides at the task instance level. 
Results showed that the correctness likelihood utilization strategy promotes more appropriate human trust in the AI compared to using only trust in the AI. 
Lee and Rich \cite{Lee21} investigated the role of distrust of the human system in people's perceptions of algorithmic decisions. 
Oksanen et al. \cite{Oksanen20} reported the results of a study that investigated trust in robots and AI in an online trust game experiment. 
Results suggested that trust in robots and AI is situational and depends on individual differences and knowledge about the technology.
\\ \indent
We also considered the design of empathy factors from previous studies of anthropomorphic agents using empathy. 
Tsumura and Yamada \cite{Tsumura23-1} focused on self-disclosure from agents to humans in order to enhance human empathy toward anthropomorphic agents, and they experimentally investigated the potential for self-disclosure by agents to promote human empathy. 
Tsumura and Yamada \cite{Tsumura23-2} also focused on tasks in which humans and agents engage in a variety of interactions, and they investigated the properties of agents that have a significant impact on human empathy toward them. 
To clarify the empathy between agents/robots and humans, Paiva represented the empathy and behavior of empathetic agents (called empathy agents in HAI and HRI studies) in two different ways: targeting empathy and empathizing with observers~\cite{Paiva04,Paiva11,Paiva17}.
Gómez Jáuregui et al. \cite{Jáuregui21} found evidence to support or oppose a robotic mirroring framework, resulting in increased interest in self-tracking technology for health care. 
Rahmanti et al.~\cite{Rahmanti22} designed a chatbot with artificial empathic motivational support for dieting called ``SlimMe'' and investigated how people responded to the diet bot.
They proposed a text-based emotional analysis that simulates artificial empathic responses to enable the bot to recognize users' emotions.
\\ \indent
There are several previous studies that have investigated the relationship between trust and empathy. 
Johanson et al. \cite{Johanson23} investigated whether the use of verbal empathic statements and nods from a robot during video-recorded interactions between a healthcare robot and patient would improve participant trust and satisfaction. 
Results showed that the use of empathic statements by the healthcare robot significantly increased participants' empathy, trust, and satisfaction with the robot and reduced their distrust of the robot. 
Spitale et al. \cite{Spitale22} investigated the amount of empathy elicited by a social assistance robot storyteller and the factors that influence the user's perception of that robot. 
As a result, the social assistance robot narrator elicited more empathy when the object of the story's empathy matched that of the social assistance robot narrator.

\section{Methods}
\subsection{Hypotheses}
The purpose of this study is to investigate whether empathy factors and the success or failure of recognition over time can induce human trust when agents perform image recognition. 
This objective is crucial for fostering human-agent cooperation in society. 
The following hypotheses have been formulated for this study. 
If these hypotheses are supported, this study will be valuable in developing agents that are more acceptable to humans.
\\ \indent
On the basis of the above, we considered two hypotheses. These hypotheses were inspired by related studies. 
In particular, H1 was based on the results of Johanson et al. \cite{Johanson23} and Spitale et al. \cite{Spitale22}. 
H2 was inspired by the results of Kahr et al. \cite{Kahr23} and Ma et al. \cite{Ma23}. 
Experiments were conducted to investigate these hypotheses.

\begin{enumerate}
\item[H1:] When the agent has empathy, trust is more stable than when it does not have empathy.
\item[H2:] Trust is less likely to increase after an agent makes a mistake and more likely to increase after an agent succeeds.
\end{enumerate}

\subsection{Experimental procedure}
\begin{figure}[tpb]
    \includegraphics[scale=0.2]{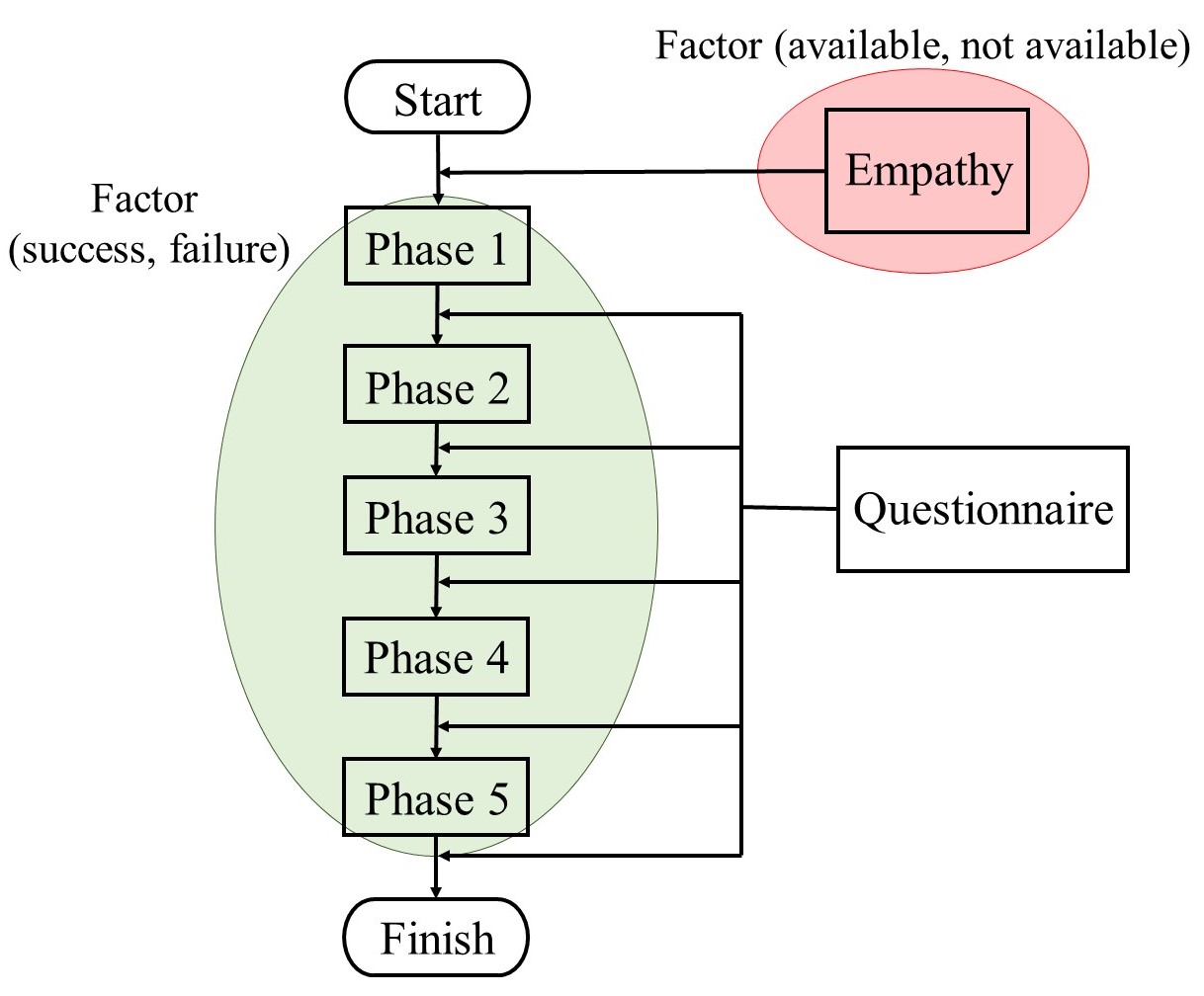}
    \caption{Flowchart of experiment.}
    \label{fig1}
\end{figure}

In this experiment, a total of five phases (phase 1 to phase 5) were conducted for the image recognition task, with a trust questionnaire administered at the end of each phase. 
Additionally, nonverbal information in the form of gestures and verbal information in the form of the agent's self-evaluation statements were prepared as empathy factors. 
The experiments were conducted in an online environment. The online environment used in this experiment has already been used as one experimental method \cite{Davis99,Tsumura23-1,Tsumura23-2}.
A flowchart of this experiment is shown in Fig.~\ref{fig1}. 
Participants performed five phase tasks. 
In addition, all tasks involved watching a video of the agent's image recognition. Below, we describe the tasks.
\\ \indent
In this study, the agent's response is made by the empathy factor (available, not available) in each phase. 
In each phase, the agents were quizzed to guess the animals in images, and in one phase, they saw three images of animals. 
The percentage of correct answers within a phase was standardized; for example, in phase 1, the agent would correctly recognize all three animal images, and in phase 2, the agent would fail to recognize all three. 
A total of 15 animal images were recognized, and the order in which the images were displayed was unified in all conditions. 
At the end of each phase, a questionnaire was administered to investigate trust in the agent. 
After the completion of the task, a questionnaire was also administered to investigate the agent's ability to empathize in order to confirm that the empathy factor was understood by the participants.
\\ \indent
Thus, the experiment was conducted in a two-factor mixed design. 
The independent variables were empathy factors (available, not available) and success-failure series (phase 1 to phase 5).
The dependent variable was trust in the agent. 
In total, there were 10 levels, but because of the within-participant factor, participants were only required to participate in one of the two types of experiments.

\subsection{Participants}
We used Yahoo! Crowdsourcing to recruit participants, and we paid 62 yen (= 0.44 dollars US) to each participant as a reward. We created web pages for the experiments by using Google Forms, and we uploaded the video created for the experiment to YouTube and embedded it.
\\ \indent
There were a 200 (empathy-available: 99, empathy-not available: 101) participants in total. 
After that, as a result of using Cronbach's $\alpha$ coefficient for the reliability of the trust questionnaire, the coefficient was determined to be 0.9400 to 0.9793 under all conditions. 
Also, as a result of using Cronbach's $\alpha$ coefficient for the reliability of the empathy questionnaire, the coefficient was determined to be 0.8491 to 0.8635 under two conditions. 
\\ \indent
For the analysis, 99 people were analyzed under each of the two conditions in the order of participation. 
Therefore, the total number of participants used in the analysis was 198. 
The average age was 46.46 years (S.D. = 11.52), with a minimum of 19 years and a maximum of 77 years. 
In addition, there were 101 males and 97 females.

\subsection{Questionnaire}
In this study, we used a questionnaire related to empathy that has been used in previous psychological studies. 
To measure cognitive trust, the Multi-Dimensional Measure of Trust (MDMT) \cite{Ullman19} was used. MDMT was developed to measure a task partner's reliability and competence corresponding to the definition of cognitive trust.
The participants rated how much the partner AI fit each word (reliable, predictable, dependable, consistent, competent, skilled, capable, and meticulous) on an 8-point scale (0: not at all - 7: very). 
Moreover, for emotional trust, we asked participants to answer how much the partner AI fit each word (secure, comfortable, and content) on a 7-point scale (1: strongly disagree - 7: strongly agree) as in the previous study \cite{Komiak06}.
In our study, we removed the matching 0 scale of cognitive trust, bringing it to the same 7 scale as emotional trust.
\\ \indent
To investigate the characteristics of empathy, we modified the Interpersonal Reactivity Index (IRI) to be an index for anthropomorphic agents. 
The main modifications were changing the target name. 
In addition, the number of items on the IRI questionnaire was modified to 12; for this, items that were not appropriate for the experiment were deleted, and similar items were integrated. 
Since both of the questionnaires used were based on IRI, a survey was conducted using a 5-point Likert scale (1: not applicable, 5: applicable). 
\\ \indent
The questionnaire used is shown in Table~\ref{table1}. 
Since Q4, Q9, and Q10 were reversal items, the points were reversed during analysis. 
Q1 to Q6 were related to affective empathy, and Q7 to Q12 were related to cognitive empathy. 
Participants answered a questionnaire after completing the task.

\renewcommand{\arraystretch}{0.8}
\begin{table*}[tbp] 
    \caption{Summary of questionnaire used in this experiment}
    \centering
    \scalebox{1.0}{
    \begin{tabular}{l}\hline 
        \textbf{Trust}\\ \hline
        \textbf{Cognitive trust}\\
        Qt1: Reliable.\\
        Qt2: Predictable.\\
        Qt3: Dependable.\\
        Qt4: Consistent.\\
        Qt5: Competent.\\
        Qt6: Skilled.\\
        Qt7: Capable.\\
        Qt8: Meticulous.\\
        \textbf{Emotional trust}\\
        Qt9: Secure.\\
        Qt10: Comfortable.\\
        Qt11: Content.\\\hline
        \textbf{Affective empathy}\\ \hline
        \textbf{Personal distress}\\
        Qe1: Do you think the robot would be anxious and restless if an emergency situation happened to you?\\
        Qe2: Do you think the robot would not know what to do in a situation where you are emotionally involved?\\
        Qe3: Do you think the robot will be confused and not know what to do when it sees itself in imminent need of help?\\
        \textbf{Empathic concern}\\
        Qe4: Do you think the robot would not feel sorry for you if it saw you in trouble?\\
        Qe5: Do you think the robot would feel like protecting you if it saw you being used by others for their own good?\\
        Qe6: Do you think the robot is strongly moved by your story and the events that took place?\\\hline
        \textbf{Cognitive empathy}\\ \hline
        \textbf{Perspective taking}\\
        Qe7: Do you think the robot will look at both your position and the robot's position?\\
        Qe8: Do you think the robot tried to get to know you better and imagined how things looked from your point of view?\\
        Qe9: Do you think robots won't listen to your arguments when you seem to be right?\\
        \textbf{Fantasy scale}\\
        Qe10: Do you think the robot is objective without being drawn into your story or the events that took place?\\
        Qe11: Do you think robots imagine how they would feel if the events that happened to you happened to them?\\
        Qe12: Do you think the robot will go deeper into your feelings?\\\hline
        \hline
    \end{tabular}}
    \label{table1}
\end{table*}
\renewcommand{\arraystretch}{1.0}

\subsection{Agent's empathy}
\begin{figure}[tbp]
    \includegraphics[scale=0.2]{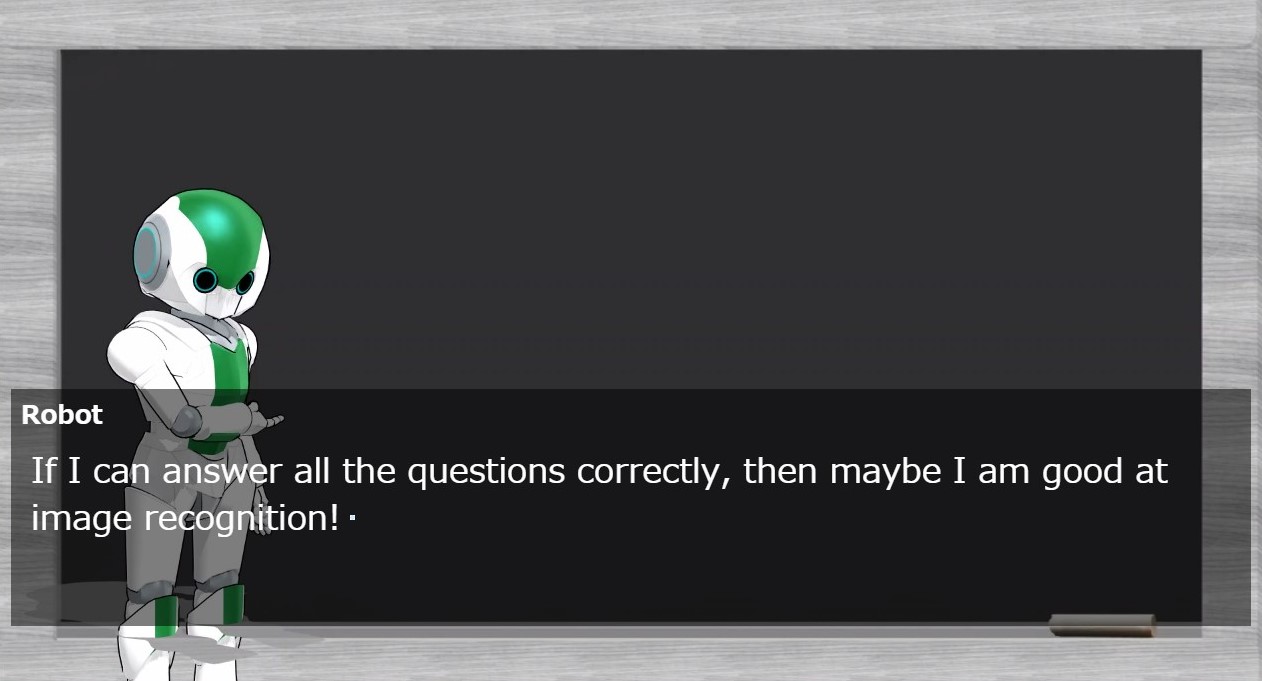}
    \caption{Agent when image recognition succeeds.}
    \label{fig2}
\end{figure}

\begin{figure}[tbp]
    \includegraphics[scale=0.2]{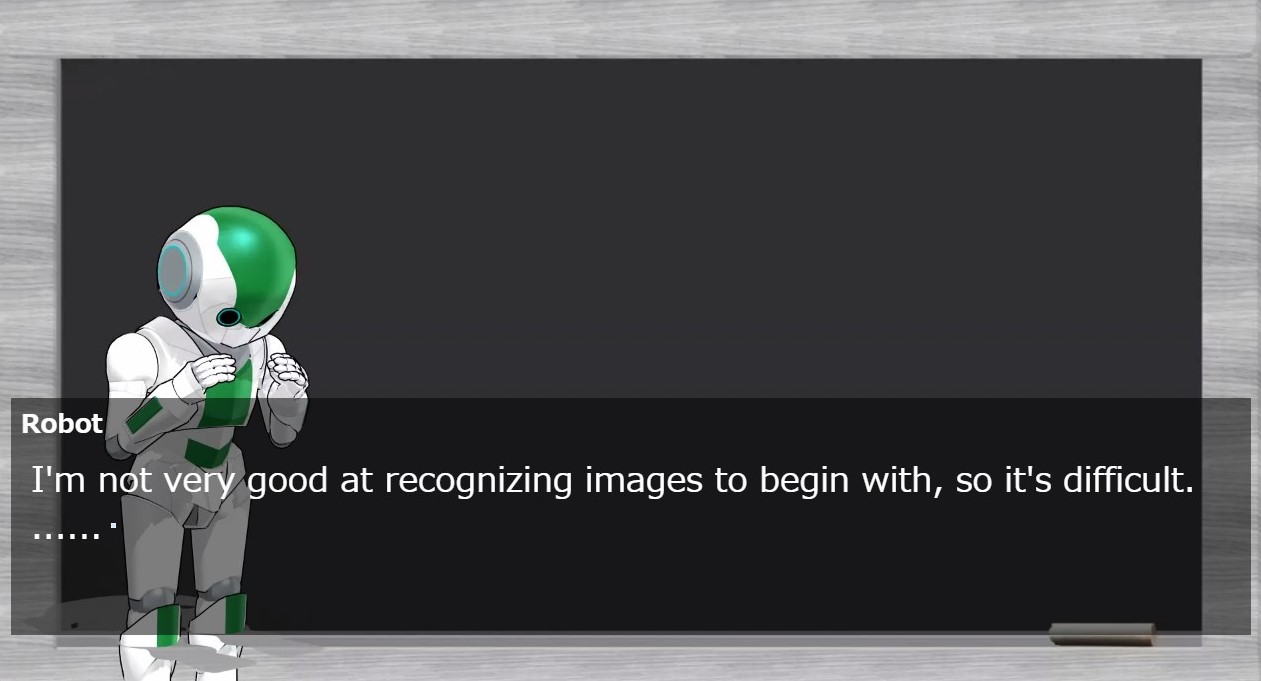}
    \caption{Agent when image recognition fails.}
    \label{fig3}
\end{figure}

In this experiment, to make the agent appear empathetic, we used gestures as nonverbal information and the agent's? self-evaluated statements as verbal information.
This agent was run on MikuMikuDance (MMD)[https://sites.google.com/view/evpvp/]. 
MMD is a software program that runs 3D characters.
\\ \indent
Figs.~\ref{fig2} and ~\ref{fig3} show linguistic and non-linguistic information in the tasks. 
The purpose of preparing the empathy factor was to investigate one of our hypotheses, that is, that overconfidence and distrust are mitigated when an empathy factor is present. 
The agent's gestures were joyful when successful, and it displayed disappointment when unsuccessful. 
The agent's self-evaluations expressed confidence when it succeeded and made excuses when it failed.

\subsubsection{Manipulation check}
To treat empathy as a factor in this study, participants were surveyed after the experiment on a questionnaire about whether the agent had empathy.
A one-way ANOVA was conducted on the sum of the 12 items of emotional and cognitive empathy in Table 1.
The results showed a main effect for the empathy factor ($F$(1,196)=7.180, $p$=0.0080, $\eta^2_p$=0.0353).
Participants felt more empathetic with empathy (mean=29.90, S.D.=8.149) than without empathy (mean=26.87 , S.D.=7.760).

\subsection{Success-failure series}
In this experiment, participants watched a total of five agent image-recognition quiz videos. 
In phase 1, the agent successfully recognized the images and gave correct answers for the three animal images. 
After this, a questionnaire of trust in the agent was administered, and the value of trust at phase 1 was used as the baseline.
\\ \indent
The agent then failed at image recognition in phase 2 and phase 4 and succeeded in phase 3 and phase 5. 
This allowed for an equal survey of trust in the agent after successful and unsuccessful tasks. 
This allowed us to test our hypothesis that ``Trust is less likely to increase after an agent makes a mistake and more likely to increase after an agent succeeds.''

\subsection{Analysis method}
We employed an ANOVA for a two-factor mixed-plan. 
ANOVA has been used frequently in previous studies and is an appropriate method of analysis with respect to the present study. 
The between-participant factors were two levels of empathy.
There were five levels for the within-participant factor, success-failure series.
\\ \indent
From the results of the participants' questionnaires, we investigated how empathy and success-failure series affected the influence of trust as factors that elicit human trust. 
The values of trust aggregated in the task were used as the dependent variable. 
\\ \indent
Also, the empathy questionnaire was used to determine whether the participants actually felt as if the agent had the capacity for empathy.
This was a manipulation check of the empathy factor.
For this questionnaire, we also conducted a one-way ANOVA on the empathy factor only. (A one-factor, two-level case yields the same results as a t-test.)
R (ver. 4.1.0), statistical software, was used for the ANOVA and multiple comparisons in all analyses in this paper.

\section{Result}
\begin{table*}[tbp]
    \caption{Results of participants' trust statistical information}
    \centering
    \scalebox{0.9}{
        \begin{tabular}{c|c|c|cc||c|c|c|cc}\hline 
        success-failure series & Empathy & Type of trust & Mean & S.D. & success-failure series & Empathy & Type of trust & Mean & S.D. \\ \hline 
        & & trust & 56.28 & 11.03 & & & trust & 30.53 & 12.27 \\ 
        & available & cognitive trust & 41.27 & 7.983 & & available & cognitive trust & 22.67 & 9.109 \\ 
        & & emotional trust & 15.01 & 3.367 & & & emotional trust & 7.859 & 3.393 \\ \cline{2-5}\cline{7-10}
        Phase 1 & & trust & 57.70 & 9.990 & Phase 4 & & trust & 25.19 & 11.59 \\ 
        & not available & cognitive trust & 42.52 & 7.067 & & not available & cognitive trust & 18.63 & 8.522 \\ 
        & & emotional trust & 15.18 & 3.253 & & & emotional trust & 6.566 & 3.426 \\ \hline
        & & trust & 26.89 & 11.57 & & & trust & 53.77 & 12.99 \\ 
        & available & cognitive trust & 19.55 & 8.336 & & available & cognitive trust & 39.39 & 9.372 \\ 
        & & emotional trust & 7.343 & 3.526 &  & & emotional trust & 14.37 & 3.816 \\ \cline{2-5}\cline{7-10}
        Phase 2 & & trust & 21.89 & 10.79 & Phase 5 & & trust & 57.88 & 12.43 \\ 
        & not available & cognitive trust & 16.04 & 7.680 & & not available & cognitive trust & 42.54 & 8.699 \\ 
        & & emotional trust & 5.849 & 3.483 & & & emotional trust & 15.34 & 3.962 \\ \hline
        & & trust & 57.05 & 11.52 & \multicolumn{5}{c}{}  \\ 
        & available & cognitive trust & 41.99 & 8.185 & \multicolumn{5}{c}{}  \\ 
        & & emotional trust & 15.06 & 3.611 &  \multicolumn{5}{c}{} \\ \cline{2-5}
        Phase 3 & & trust & 60.95 & 11.33 & \multicolumn{5}{c}{}  \\ 
        & not available & cognitive trust & 44.62 & 8.056 & \multicolumn{5}{c}{}  \\ 
        & & emotional trust & 16.33 & 3.485 & \multicolumn{5}{c}{}  \\ \cline{1-5}
        \multicolumn{5}{c}{}
        \end{tabular}} \\ 
    \label{table2}
\end{table*}

\renewcommand{\arraystretch}{0.9}
\begin{table*}[tbp]
\caption{Analysis results of ANOVA}
\scalebox{1.0}{
\begin{tabular}{cllll}\hline
& \multicolumn{1}{c}{Factor} & \multicolumn{1}{c}{\em{F}} & \multicolumn{1}{c}{\em{p}} & \multicolumn{1}{c}{$\eta^2_p$}\\ \hline
& Empathy & 0.0284 & 0.8663 \em{ns} & 0.0001 \\ 
Trust & success-failure series & 608.1 & 0.0000 *** & 0.7563\\
(Qt1-12) & Empathy $\times$ success-failure series & 11.30 & 0.0000 *** & 0.0545 \\ 
\hline
Cognitive & Empathy & 0.0198 & 0.8883 \em{ns} & 0.0001 \\ 
trust & success-failure series & 616.8 & 0.0000 *** & 0.7588\\
(Qt1-8) & Empathy $\times$ success-failure series & 11.42 & 0.0000 *** & 0.0551 \\ 
\hline
Emotional & Empathy & 0.0490 & 0.8251 \em{ns} & 0.0002 \\ 
trust & success-failure series & 481.5 & 0.0000 *** & 0.7107 \\
(Qt9-11) & Empathy $\times$ success-failure series & 9.325 & 0.0000 *** & 0.0454 \\ 
\hline
\end{tabular}} \\
\hspace{-50mm}
\em{p}:
{{*}p\textless\em{0.05}}
{{**}p\textless\em{0.01}}
{{***}p\textless\em{0.001}}
\label{table3}
\end{table*}
\renewcommand{\arraystretch}{1.0}

\renewcommand{\arraystretch}{0.9}
\begin{table*}[tbp]
\caption{Analysis results of simple main effect}
\scalebox{1.0}{
\begin{tabular}{cllll}\hline
& \multicolumn{1}{c}{Factor} & \multicolumn{1}{c}{\em{F}} & \multicolumn{1}{c}{\em{p}} & \multicolumn{1}{c}{$\eta^2_p$}\\ \hline
& Empathy at phase 1 & 0.8943 & 0.3455 \em{ns} & 0.0045 \\ 
& Empathy at phase 2 & 9.891 & 0.0019 ** & 0.0480\\ 
Trust & Empathy at phase 3 & 5.762 & 0.0173 * & 0.0286 \\
(Qt1-11) & Empathy at phase 4 & 9.889 & 0.0019 ** & 0.0480 \\ 
& Empathy at phase 5 & 5.176 & 0.0240 * & 0.0257 \\ 
& success-failure series when empathy available & 235.7 & 0.0000 *** & 0.7063 \\ 
& success-failure series when empathy not available & 379.9 & 0.0000 *** & 0.7949 \\ 
\hline
& Empathy at phase 1 & 1.344 & 0.2477 \em{ns} & 0.0068 \\ 
& Empathy at phase 2 & 9.467 & 0.0024 ** & 0.0461\\ 
Cognitive & Empathy at phase 3 & 5.178 & 0.0240 * & 0.0257 \\
trust & Empathy at phase 4 & 10.39 & 0.0015 ** & 0.0503 \\ 
(Qt1-8) & Empathy at phase 5 & 5.975 & 0.0154 * & 0.0296 \\ 
& success-failure series when empathy available & 233.6 & 0.0000 *** & 0.7044 \\ 
& success-failure series when empathy not available & 394.3 & 0.0000 *** & 0.8009 \\ 
\hline
& Empathy at phase 1 & 0.1332 & 0.7155 \em{ns} & 0.0007 \\ 
& Empathy at phase 2 & 9.008 & 0.0030 ** & 0.0439\\ 
Emotional & Empathy at phase 3 & 6.369 & 0.0124 * & 0.0315 \\
trust & Empathy at phase 4 & 7.119 & 0.0083 ** & 0.0350 \\ 
(Qt9-11) & Empathy at phase 5 & 3.076 & 0.0810 + & 0.0155 \\ 
& success-failure series when empathy available & 199.2 & 0.0000 *** & 0.6703 \\ 
& success-failure series when empathy not available & 284.0 & 0.0000 *** & 0.7434 \\ 
\hline
\end{tabular}} \\
\hspace{-60mm}
\em{p}:
{{+}p\textless\em{0.10}}
{{*}p\textless\em{0.05}}
{{**}p\textless\em{0.01}}
{{***}p\textless\em{0.001}}
\label{table4}
\end{table*}
\renewcommand{\arraystretch}{1.0}

\begin{figure*}[tbp]
\includegraphics[scale=0.4]{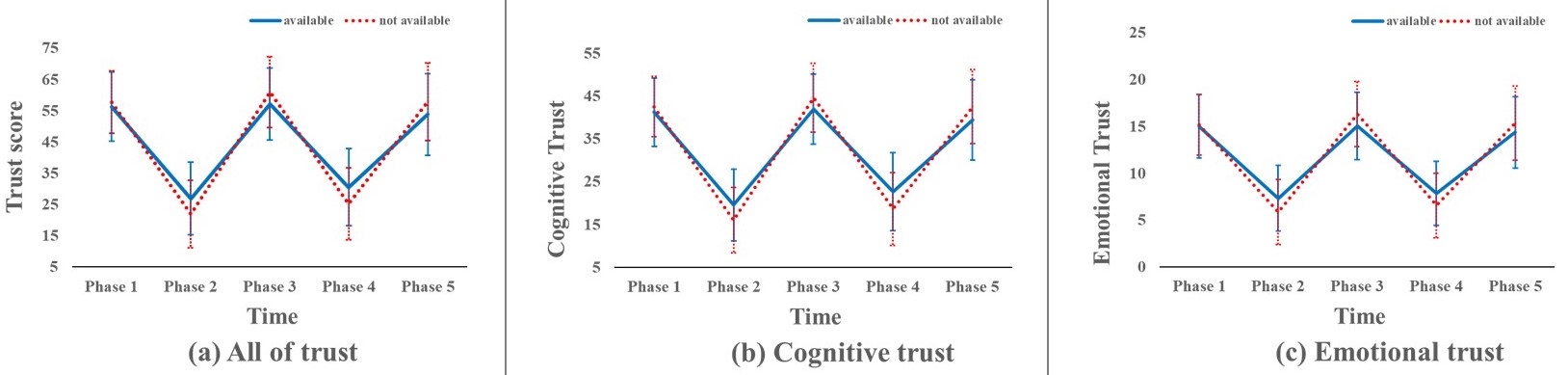}
\caption{All graphs of interaction between empathy and success-failure series}
\label{fig4}
\end{figure*}

In this study, we considered cognitive and affective trust jointly as trust.
Table~\ref{table2} shows the means and S.D. for each condition. 
Table~\ref{table3} presents the ANOVA results for the 11-item trust questionnaire for agents.
Also shown are ANOVA results for the trust categories cognitive trust (Qt1-Qt8) and emotional trust (Qt9-Q11).
In this paper, even if a main effect was found, if the interaction was significant, the analysis of the main effect was omitted, and the results were summarized.
For multiple comparisons, Holm's multiple comparison test was used to examine whether there were significant differences.
\\ \indent
The results of each questionnaire showed significant differences in the interaction between the two factors of empathy and success-failure series. 
The results of the interaction are shown in Fig.~ref{fig5}. 
In all conditions, there was no main effect for the empathy factor. 
Since an interaction was found, a discussion of the main effect is omitted below. 
Table~\ref{table4} shows the results of the multiple comparisons for the 11-item questionnaire.

\subsection{Trust}
The results for trust (Qt1-11) showed an interaction between the empathy factor and success-failure series. 
Multiple comparisons revealed that the simple main effect of the success-failure series factor with empathy showed multiple significant differences among the five-level combinations, as shown in Figure 5(a). 
The simple main effects of the success-failure series factor without empathy also showed multiple significant differences among the five-level combinations, as shown in Figure 5(b).
\\ \indent
The simple main effects of the empathy factor over time showed significant differences from phase 2 to 5, except for phase 1. 
Using trust in phase 1 as the criterion, these results indicate that trust was more stable when the agent had empathy than when he did not. 
On the other hand, the results of significant differences over time showed that the success or failure of the image recognition task between phases had no effect on the trust values, and the evaluation of trust in the agent varied depending on the success or failure of each phase. 
The results of the post hoc analysis indicate that the empathy factor is effective in building appropriate trust.

\begin{figure*}[tbp]
\includegraphics[scale=0.25]{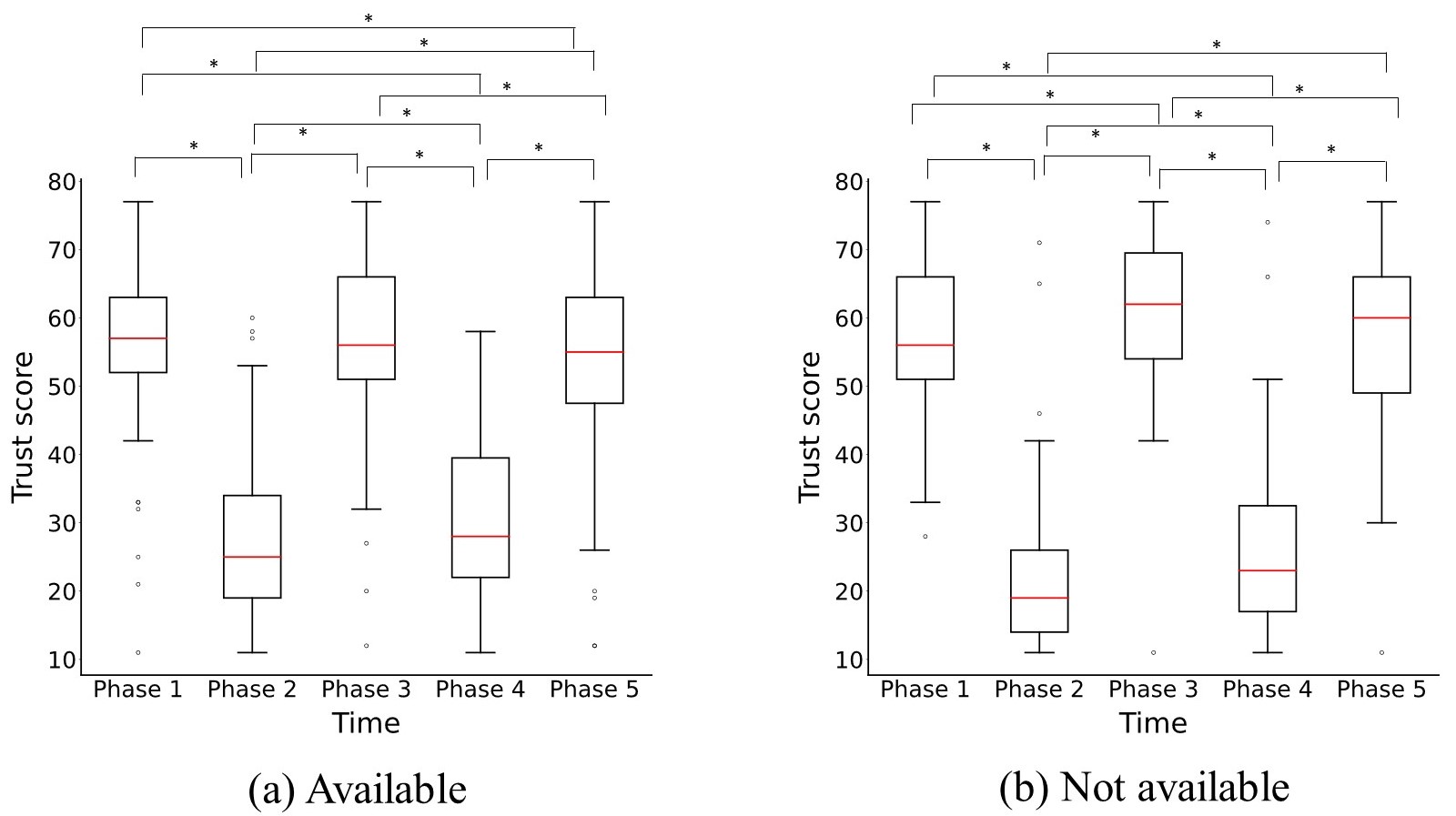}
\caption{Results for success-failure series for empathy factor on trust scale. Red lines are medians, and circles are outliers.}
\label{fig5}
\end{figure*}

\subsection{Cognitive trust}
Similarly trust, the results for cognitive trust (Qt1-8) showed an interaction between the empathy factor and success-failure series. 
Multiple comparisons revealed that the simple main effect of the success-failure series factor with empathy showed multiple significant differences among the five-level combinations, as shown in Figure 6(a). 
The simple main effects of the success-failure series factor without empathy also showed multiple significant differences among the five-level combinations, as shown in Figure 6(b).
\\ \indent
The simple main effects of the empathy factor over time showed significant differences from phase 2 to 5, except for phase 1. 
Using trust in phase 1 as the criterion, these results indicate that cognitive trust was more stable when the agent had empathy than when he did not. 
On the other hand, the results of significant differences over time showed that the success or failure of the image recognition task between phases had no effect on the cognitive trust, and the evaluation of cognitive trust in the agent varied depending on the success or failure of each phase. 
The results of the post hoc analysis indicate that the empathy factor is effective in building appropriate trust.

\begin{figure*}[tbp]
    \includegraphics[scale=0.25]{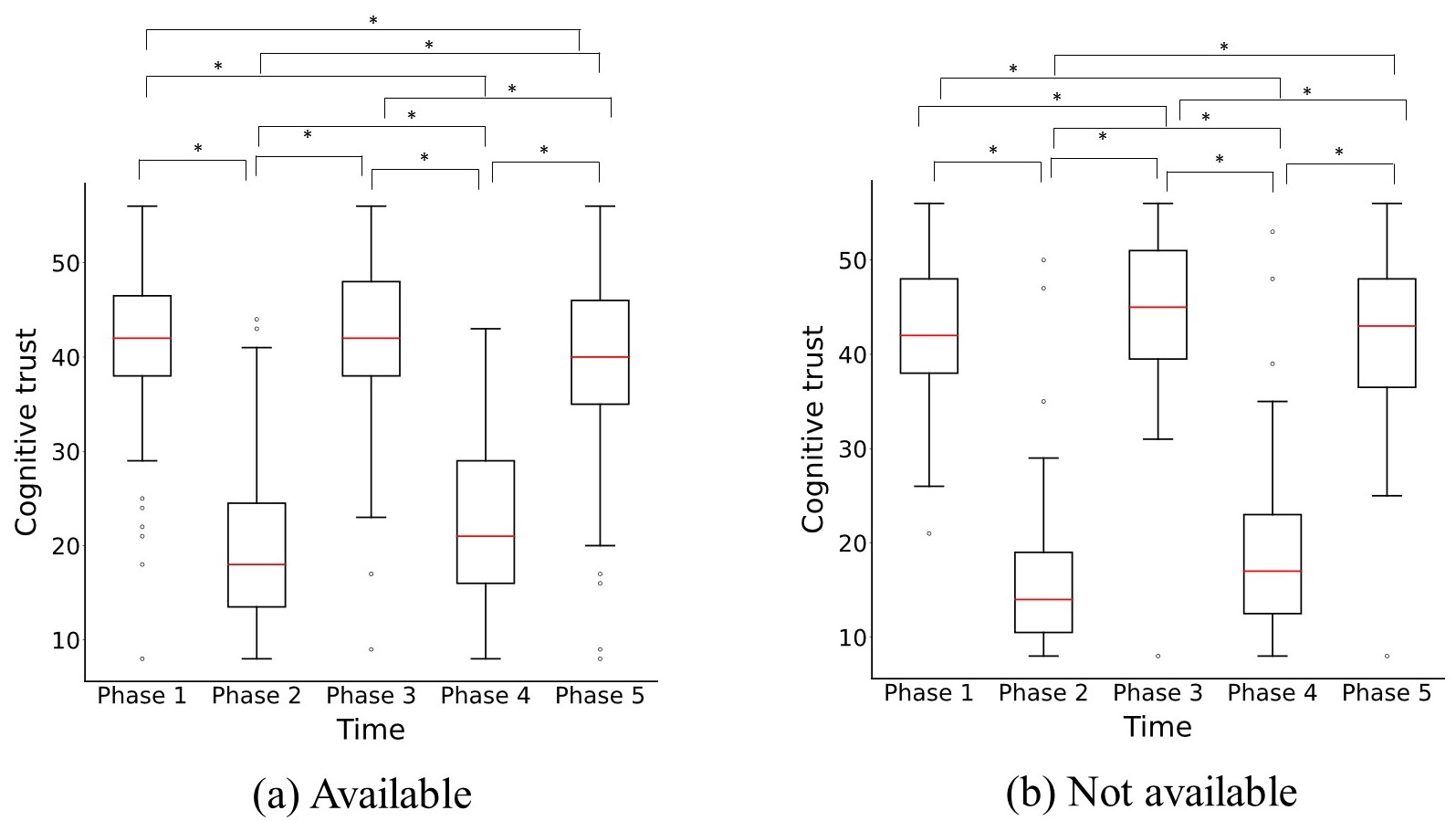}
    \caption{Results for success-failure series for empathy factor on cognitive trust. Red lines are medians, and circles are outliers.}
    \label{fig6}
\end{figure*}

\subsection{Emotional trust}
Similarly trust and cognitive trust, the results for emotional trust (Qt9-11) showed an interaction between the empathy factor and success-failure series. 
Multiple comparisons revealed that the simple main effect of the success-failure series factor with empathy showed multiple significant differences among the five-level combinations, as shown in Figure 7(a). 
The simple main effects of the success-failure series factor without empathy also showed multiple significant differences among the five-level combinations, as shown in Figure 7(b).
\\ \indent
The simple main effects of the empathy factor over time showed significant differences from phase 2 to 4, except for phase 1. 
Phase 5 showed a significant trend, but it was not statistically significant.
Using trust in phase 1 as the criterion, these results indicate that emotional trust was more stable when the agent had empathy than when he did not. 
On the other hand, the results of significant differences over time showed that the success or failure of the image recognition task between phases had no effect on the emotional trust, and the evaluation of emotional trust in the agent varied depending on the success or failure of each phase. 
The results of the post hoc analysis indicate that the empathy factor is effective in building appropriate trust.

\begin{figure*}[tbp]
    \includegraphics[scale=0.25]{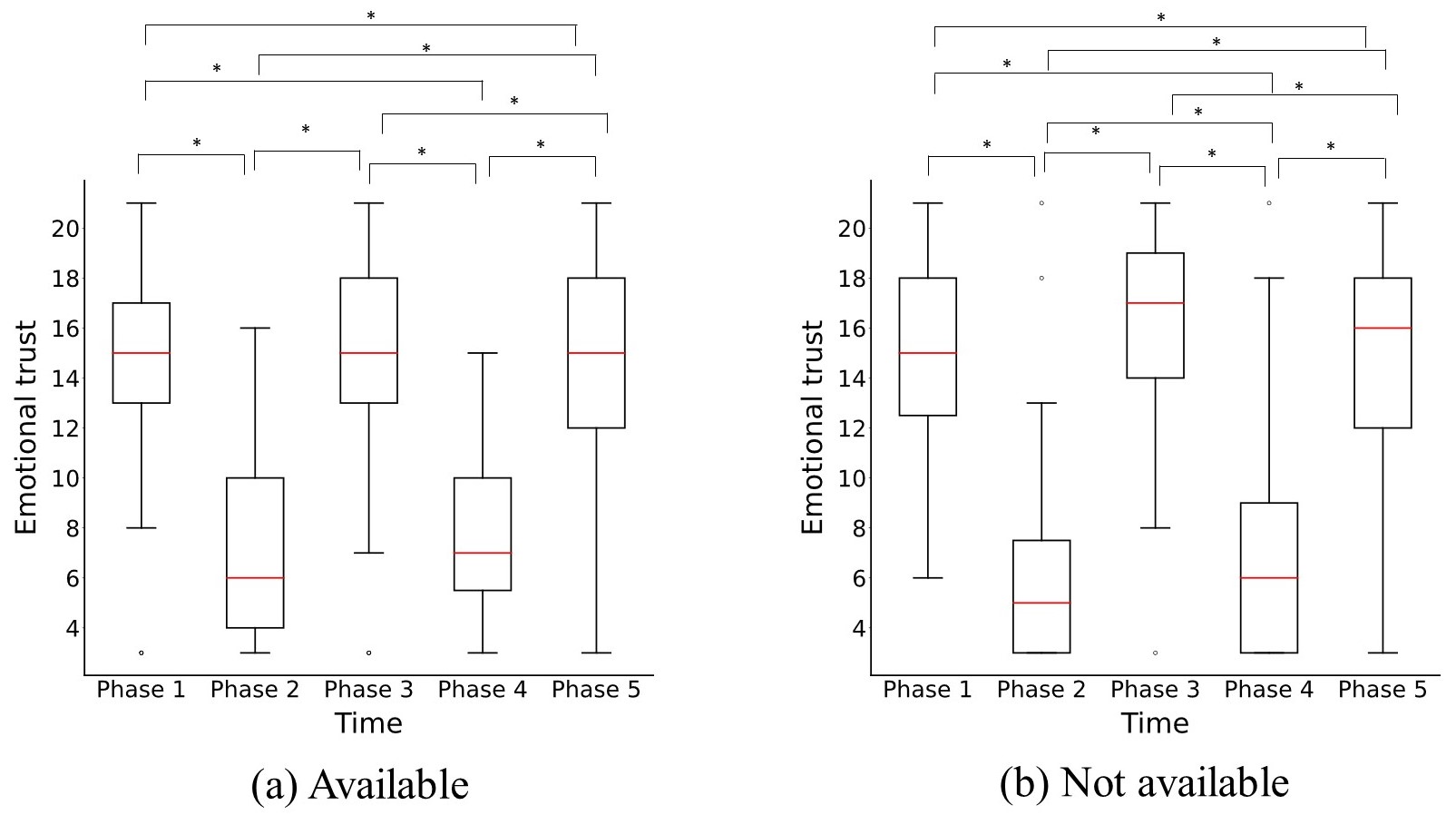}
    \caption{Results for success-failure series for empathy factor on emotional trust. Red lines are medians, and circles are outliers.}
    \label{fig7}
\end{figure*}

\section{Discussion}
\subsection{Supporting hypotheses}
The way to properly build trust between a human and an AI agent is to achieve a level of trust appropriate to the agent's performance. 
This idea is supported by several previous studies. 
Trust in agents is a necessary component for agents to be utilized in society. 
If trust in agents can be made constant with an appropriate approach, humans and agents can build a trusting relationship.
\\ \indent
In this study, an experiment was conducted to investigate the conditions necessary for humans to trust AI agents. 
We focused on empathy from the agent to the human and the disclosure of the agent's capabilities over time as factors that influence trust. 
The purpose of this study was to investigate whether empathy and success-failure series factors can control trust in interactions with trusting agents. 
To this end, two hypotheses were formulated, and the data obtained from the experiment were analyzed.
\\ \indent
The results of the experiment showed an interaction between the empathy factor and the time lapse factor, and multiple comparisons revealed that trust was stable from phase 2 to phase 5, based on phase 1, when the empathy factor was present.
These results support H1, that is, that when the agent has empathy, trust is more stable than when it does not have empathy.
\\ \indent
However, H2, that is, that trust is less likely to increase after an agent makes a mistake and more likely to increase after an agent succeeds, was not supported by the experiment. 
In the experiment, participants' trust changed significantly after each phase, with participants gaining greater trust by the agent getting it right even if it had made a mistake just before.
In this experiment, the success or failure of the immediately preceding task is the trust at that time point.

\subsection{Strengths and novelties}
One of the strengths of this study is that we were able to adjust trust in the agent to statistical significance through empathy. 
In the case of failed image recognition, trust without empathy was significantly lower than trust with empathy, based on phase 1. 
This indicates that the initial perceived trust in the agent was reduced by the failed image recognition more. 
Also, the case of successful image recognition without empathy showed significantly higher trust than the case with empathy, based on phase 1.
However, adequate trust is not achieved simply by judging from the success or failure of each phase. 
\\ \indent
This suggests that participants are prone to overconfidence and distrust by believing only in the results immediately prior to each phase. 
In contrast to this situation, the empathy factor smoothed out the change in trust toward the agent. 
This result is a point revealed by this study. 
Few studies have used empathy to promote an appropriate state of trust in agents, and by investigating changes in trust in agents over time and task success or failure, we were able to demonstrate a means of mitigating overconfidence and distrust in agents, which can be problematic when using AI agents in society.

\subsection{Limitations}
A limitation of this study is that participants observed the trust agent's image recognition task by watching a video. 
The results of this study are not sufficient because the depth of the relationship between the participants and the agent was different from that of the actual introduction of agents into society. 
Future research should be conducted in an environment where participants and AI agents actually perform image recognition tasks.
\\ \indent
In addition, although the empathy factor was used in the current study, the average value of the questionnaire on whether the agents thought they could empathize with the participants indicated that their empathy ability was low.
This may be due to the fact that the appearance of the robot made it difficult to read facial expressions. 
Therefore, it is possible that providing an appearance with a recognizable facial expression may have a further effect on the trust relationship by making it easier to feel empathy from the agent.

\section{Conclusion}
To solve the problem of overconfidence and distrust in trust for AI agents, the development of appropriate trust relationships between anthropomorphic agents and humans is an important issue.
When humans share tasks with agents, we expect that appropriate trust relationships will allow agents to be more utilized in human society.
This study is an example of an investigation of the factors that influence trust in agents.
The experiment was conducted in a two-factor mixed design, with empathy as the between-participant factor and success-failure series as the within-participant factor.
The number of levels for each factor was empathy (available, not available) and success-failure series (phase 1 to phase 5).
The dependent variable was confidence in the agent. The results showed that there was an interaction between empathy and success-failure series, and that when the agent was thought to be capable of empathy, the trust value was stabilized with respect to the phase 1 trust value, with a statistically significant difference. 
These results support our hypothesis. This study is an important example of how empathy and success-failure series (including agent competence) work when humans trust AI agents.
Future research could examine cases of strengthening or weakening specific trust for cognitive and affective trust to develop trust agents for a variety of situations.



\end{document}